\documentclass[10pt]{aastex}
\usepackage{graphicx}

\usepackage{emulateapj5}
\usepackage{apjfonts}
 
\shorttitle{Interferometry and asteroseismology}
\shortauthors{Creevey et al.}

\begin{document}
\title{The complementary roles of interferometry 
and asteroseismology in determining the mass
of solar-type stars}


\author{
O.~L.~Creevey\altaffilmark{1,2a,3},
M.~J.~P.~F.~G. Monteiro\altaffilmark{2,4},
T.~S.~Metcalfe\altaffilmark{1},\\
T.~M.~Brown\altaffilmark{1,5},
S.~J.~Jim\'enez-Reyes\altaffilmark{3,6},
J.~A.~Belmonte\altaffilmark{3}
}

\email{creevey@hao.ucar.edu}
\email{mjm@astro.up.pt}
\email{tbrown@lcogt.net}
\email{travis@hao.ucar.edu}
\email{jba@iac.es}
\email{sjimenez@iac.es}

\altaffiltext{1}{High Altitude Observatory, National Center for 
Atmospheric Research, Boulder, Colorado 80301, USA.}
\altaffiltext{2}{Centro de Astrof\'{\i}sica da Universidade do Porto, Rua 
das Estrelas 4150, Porto, Portugal.}
\altaffiltext{3}
{Instituto de Astrof\'{\i}sica de Canarias, 
C/ V\'ia L\'actea S/N, Tenerife, Spain E-38200.}
\altaffiltext{4}{Departamento de Matem\'atica Aplicada, 
Faculdade de Ci\'encias da Universidade do Porto, Portugal}
\altaffiltext{5}{present address: Las Cumbres Observatory Inc, 
6740 Cortona Dr. Goleta, California 93117, USA.}
\altaffiltext{6}{present address: School of Physics and Astronomy, University
of Birmingham, UK}
\altaffiltext{2}{\hspace{-0.1cm}$^{\rm A}$Visiting Scientist}

\begin{abstract}
How important is an independent diameter measurement
for the determination of stellar parameters of 
solar-type stars? 
When coupled with seismic observables, how well
can we determine the stellar mass?
If we can determine the radius of the star to between
1\% and 4\%, how does this affect the theoretical
uncertainties?
Interferometry can provide an independent radius determination
and it has been suggested that we should expect
at least a 4\% precision on such a measurement for
nearby solar-type stars.
This study aims to provide both qualitative and 
quantitive answers to these questions for a
star such as our Sun, where seismic information
will be available.
We show that the importance of an independent
radius measurement depends on the combination of 
observables available {\it and} the size of 
the measurement errors.
It is important for determining all stellar
parameters and in particular the mass, where
a good radius measurement
can even allow us to determine the mass with a precision better than 2\%.
Our results also show that measuring the small frequency
separation $\delta \nu$ significantly 
improves the determination of 
the evolutionary stage $\tau$ and the mixing-length parameter $\alpha$.
\end{abstract}



\section{Introduction}

Asteroseismology is the interpretation of a star's
oscillation frequency spectrum to characterize
its internal structure. 
By probing its interior, we are testing our
knowledge of fundamental physics.
A star with oscillations such as our Sun (solar-type star)
presents a frequency spectrum with a range
of excited modes.
Because most of the proper modes are excited,
mode identification for each frequency is considerably
easier than for other types of pulsating stars \citep{kje04,bed04,kje05}.

Modelling these stars can be difficult
as there are often multiple sets of parameters
that fit the observations.
Also, if we use the frequencies to determine the 
mass of the star, then any solution we find
may be incorrect as it depends on the {\it fit} mass.
If we could determine the mass of the star through
some other means, then modelling the
frequencies would be easier, {\it and}
they could also then be used solely to characterize the
internal structure of the star.

Unfortunately the precise determination of
mass is normally possible only for detached components of similar
mass in
spectroscopic binary systems. 
For single stars we use observables such as magnitudes and 
colors 
(apart from oscillation frequencies)
to estimate the mass. However,
now that there are various possibilites
of measuring a radius for single stars through interferometry 
\citep{par03,gli03,ker03,auf06} the question then arises:
Can radius be used to determine the mass of the star?

Recent studies have shown interferometry has
determined
the radius of a star
with a precision of about 1\% for the 
brightest stars and it has been suggested
that we might obtain a precision of at least 4\% for 
most nearby solar-type stars whose diameters
we could expect to measure \citep{pij03,the05,dif04}.
Our interest is exploring how 
this independent measure 
complements oscillation frequency information for solar-type stars.
Can it determine the mass of the
star?
And if so, can we then use the frequencies to 
probe the physics of the star?
How important is the radius for determining other
stellar parameters such as age and chemical composition?
What is the impact on the expected uncertainties? 
Now that we are expecting frequency errors 
less than about 1.0$\mu$Hz with the launch of 
CoRoT\footnote{http://smsc.cnes.fr/COROT} \citep{bag00}
and 
Kepler\footnote{http://kepler.nasa.gov} \citep{bor97,bor04,bas04}, 
does this have an effect on our results?

While the objective of this study is mainly to determine
stellar mass, 
we also discuss results for other stellar parameters.
Section~\ref{sec-background} describes the mathematical background, 
the physics of the models and the observables we use for
this study.
Section~\ref{sec-theoretical} discusses some theoretical
results indicating the importance of a radius measurement.
Section~\ref{sec-simulations} presents results of 
simulations which support the theoretical predictions.

\section{Description of the Problem
\label{sec-background}}

\subsection{Mathematical Solution}
The techniques incorporated for this study follow
that of \citet{bro94a} and \citet{mig05}.
For a more elaborate description we refer the readers to 
\citet{bro94a} or \citet{num92} and summarize here the main concepts for the
purposes of understanding our work.

\subsubsection{Finding the parameters}
Taylor's Theorem allows us to approximate any differentiable
function near a point {\bf x$_0$} by a polynomial that depends
only on the derivatives of the function at that point.
To first order this can be written as
\begin{equation}
  {\bf f}({\bf x}) = {\bf f}({\bf x_0}) + 
  {\bf f'}({\bf x_0})({\bf x} - {\bf x_0})
  \label{eqn-taylor}
\end{equation} 
where  
${\bf x} = \{x_j\}_{j=1}^N$ are the $N$ parameters defining 
the system (``input parameters'') and
${\bf f} = \{f_i\}_{i=1}^M$ are the $M$ expected outputs of
the system that depend on ${\bf x}$.
These are the expected measurements or {\it observables}.
To distinguish between 
the expected observables and real 
observations we denote the latter by
${\bf O} =  \{O_i\}_{i=1}^M$.

We measure ${\bf f}$ 
and would like to find the set ${\bf x}$ that produce these 
measurements.  
This problem is a typical inverse problem whose
solution ${\bf x}$ (in the pure linear case) falls neatly out 
from 
\begin{equation}
{\bf x} = {\bf x_0} + {\bf V W ^{-1} U^T} {\bf \delta f},
\label{eqn-solution}
\end{equation}
where 
\begin{equation}
{\bf \delta f} = \frac{{\bf O} - {\bf f}({\bf x_0})}{{\bf \epsilon}}.
\label{eqn-deltao}
\end{equation}
Here
\begin{equation}
{\bf f'}({\bf x_0}){\bf \epsilon^{-1}} = {\bf D} = {\bf UWV^T}
\label{eqn-svd}
\end{equation} 
is the {\it Singular Value Decomposition} (SVD) of the 
derivative matrix (section \ref{sec-math2})
and
${\bf \epsilon} = \{\epsilon_i\}_{i=1}^M$ 
are the measurement errors.

In the case of a non-linear physical model, 
the method for finding the true values ${\bf x}$
is to use a modified version of equation (\ref{eqn-solution}) iteratively.
Incorporating
a goodness-of-fit test such as a $\chi^2$ function
\begin{equation}
\chi^2 = \sum_{i=1}^M \frac{(f_i({\bf x}) - O_i)^2}{\epsilon_i^2}
\label{eqn-chi2}
\end{equation}
allows us to find
the best set of parameters ${\bf x}$ by minimizing $\chi^2$.

There are many reliable algorithms available for
performing such minimizations \citep{num92}.
The algorithm implemented in this work is the 
{\it Levenberg-Marquardt} algorithm
because it is known to be robust and incorporates derivative 
information.  
It also usually converges within 2-3 iterations
to its minimum.

\subsubsection{Singular Value Decomposition  \label{sec-math2}}
SVD is the factorization of any $M\times N$ matrix ${\bf D}$
into 3 components ${\bf U}$, ${\bf V^T}$ 
and ${\bf W}$ (eq.~(\ref{eqn-svd})).
${\bf V^T}$ is the transpose of ${\bf V}$ which is an 
$N \times N$ orthogonal matrix that contains the
{\it input} basis vectors for ${\bf D}$, or the vectors
associated with the parameter space.
${\bf U}$ is an $M \times N$ orthogonal matrix that contains the
{\it output} basis vectors for ${\bf D}$, or the vectors
associated with the observable space.
${\bf W}$ is a diagonal matrix that contains the
{\it singular values} of ${\bf D}$.  

We use SVD in our analysis because 
it provides a method to investigate
the information content of our observables
and their impact on each of the parameters.
Equations (\ref{eqn-deltao}) and (\ref{eqn-svd}) were
defined in terms of the observed or expected error so
SVD can be used to study various properties of 
{\bf D}:
\begin{itemize}
\item The expected uncertainties
in each of the 
parameters via the covariance matrix
\begin{equation}
C_{jk} = \sum_{i=1}^N \frac{V_{ji}V_{ki}}{W_{ii}^2},
\label{eqn-covariance}
\end{equation}
\item The significance of each of the observables
for the determination of the parameter solution
\begin{equation}
S_i = \left (  \sum_{j=1}^N U_{ij}^2 \right )^{1/2}.
\label{eqn-significance}
\end{equation}
\end{itemize}

\subsection{Parameters \& Observables}

For this study we need to distinguish clearly
between our parameters and observables. 
The {\it parameters} are the input ingredients ${\bf x}$
to our physical
model.
For example in a stellar code the parameters
include mass, age and chemical composition.
The {\it observables} ${\bf f}$ are then the outputs 
of the models
given these ${\bf x}$ and are those things
that we can normally measure, such as magnitudes, 
effective temperature and oscillation frequencies.
To refrain from ambiguity between the observable
errors and the uncertainties in the parameters
we shall denote the former by $\epsilon$ and
the latter by $\sigma$.

\subsubsection{Parameters \& Models}
We describe a solar-type star with five adjustable parameters.
These are mass $M$, age (or evolutionary stage) $\tau$,
chemical composition given by two of $(X,Y,Z)$ where $X+Y+Z=1$
and $X$, $Y$ and $Z$ are the initial hydrogen, helium
and metal abundance respectively, and
the mixing-length parameter $\alpha$ which describes 
convection in the outer envelope of the star.

We use the {\it Aarhus STellar Evolution Code} (ASTEC)
for stellar structure and evolution and the 
{\it ADIabatic PuLSation} code (ADIPLS) to calculate
non-radial oscillation frequencies \citep{chr82}.
For each model evolved with the parameters  
$(M,\tau,X,Z,\alpha)$ ASTEC produces
stellar quantities 
which are then used 
to calculate the oscillation frequencies with ADIPLS.
The outputs from the models implemented
in this work are the global structure parameters
such as radius $R_{\star}$, effective temperature $T_{\rm eff}$, 
and a set of oscillation frequencies for $l=0,1,2,3$ and
$n=1,2,...,32$.
The Basel model atmospheres \citep{lej97} are used to calculate 
magnitudes and colors.
$R_{\star}$, $T_{\rm eff}$, $\log g$, and $[Z/X]$ are used 
to obtain spectra which are then integrated in a range
of wavelengths to produce magnitudes.

We fixed the physics of the stellar evolution models.
The equation of state (EOS) is that of \citet{egg73}.
The opacities are the OPAL 1995 tables \citep{rog96},
supplemented
by Kurucz opacities at low temperatures.
Convection is described by the classical {\it mixing-length
theory} \citep{boh58} where the mixing-length $\ell$ is defined as
$\ell=\alpha H_p$, $\alpha$ is the free {\it mixing-length parameter}
and $H_p$ is the local pressure scale height.
Diffusion is not included and so we allow 
$[$M/H$]\sim[$Fe/H$]$ to estimate $[Z/X]$
where we are assuming a near solar-composition.
We ignore overshoot effects
and do not include Coulomb corrections.
This study concentrates on one main-sequence model star 
with solar-type oscillations whose 
parameters are given in Table~\ref{tbl-parameters} but
is extendable to a solar-type star with similar 
characteristics.

\subsubsection{Observables \label{sec-observables}}

Typical observables for such a star come from 
various sources of observations.
Spectroscopy provides  
effective temperature $T_{\rm eff}$,
gravity $\log g$ and
metallicity $[$M/H$]$ measurements.
With photometry we obtain
magnitudes e.g. $V$ and
colors e.g. $(U-V)$.
Oscillation frequencies are obtained using time-series
observations of photometric and/or spectroscopic origin.
Interferometry measures a 
limb-darkened\footnote{Information regarding the light intensity
profile across the star is provided by the interferometric
observations of the {\it second lobe of the 
visibility function}.  In the case where the star is not fully resolved,
i.e. we only get some visibility points on the first lobe, then
we require a limb-darkened model for the star, rendering $R_{\star}$ 
model-dependent.  The limb-darkened profile also varies if we use
1-D or 3-D atmospheric models.  Both \citet{auf05} and \citet{big06} provide
detailed and quantitative descriptions on this matter 
and also confirm the inadequacy of the 1-D models in
reproducing these second-lobe measurements.} 
stellar angular diameter in milliarcseconds (mas) 
and 
this coupled with a parallax gives an absolute diameter
value (we use the radius $R_{\star}$ value for this work).

The observables $O_i$ and expected errors $\epsilon_i$ 
for our model star 
are given in 
Table~\ref{tbl-observables}.
The measurement errors reflect what is most currently quoted in the
literature.  However we do realize that some of these values are optimistic,
and are unlikely to be improved upon.  In this way, we can investigate
how much further interferometry and asteroseismology can take us for
determining stellar mass.

We used frequency separations $\Delta \nu_{\rm n,l}$ and $\delta \nu_{\rm n,0}$
instead of individual frequencies
$\nu_{\rm n,l}$ where
\begin{equation}
\Delta \nu_{\rm n,l} = \nu_{\rm n,l} - \nu_{\rm n-1,l},
\label{eqn-largeseparation}
\end{equation}
\begin{equation}
\delta \nu_{\rm n,0} = \nu_{\rm n+1,0} - \nu_{\rm n,2}.
\label{eqn-smallseparation}
\end{equation}
While we realize that we are throwing away important
information, the reasons for doing this 
are justified: the high sensitivity of 
the frequency values to changes in the parameters
(non-linearity of the problem), and
the unreliability of the treatment of the stellar surface
layers and hence absolute values of $\nu_{\rm n,l}$.

We also assume that we can identify the $(l,n)$ quantum
numbers.  Rotation will induce a small effect (solar-type star) 
so we can assume
$(l,m) \sim (l,m=0)$.
The degree $l$ can be interpreted by using
echelle-diagrams (see Figure 6 in \citet{kje05}).  
It is only the radial order $n$ that
is then difficult to identify.  
This number is 
model-dependant, and there is no direct way of {\it observing} it.
However with a rich frequency spectrum expected of these stars,
the relative $n$ position of each mode
of degree $l$ can be identified and then also
the average frequency separations
$\bar{\Delta \nu}$ and 
$\bar{\delta \nu}$ (independent of $n$).
These observables are used for the initial fitting process, and
an inspection of
the $\nu_{\rm n,l}$ that come from the models allows an 
$n$ identification to $\pm$1.

\section{Various Roles of the Observables \label{sec-theoretical}}
Why should a diameter measurement such as that expected
from interferometry be important?
This section highlights the importance that a radius measurement 
has when coupled with various observables
and different measurement errors.

\subsection{Significance}
The observables play very different roles in the determination
of the parameter solution depending on which 
combination of observables ($O_i$) are available and the size
of their measurement errors ($\epsilon_i$).
Using equation (\ref{eqn-significance}) we calculated the 
significance of some observables $S(O_i)$ using different
observable combinations.

\begin{figure}
\includegraphics[width = 9cm]{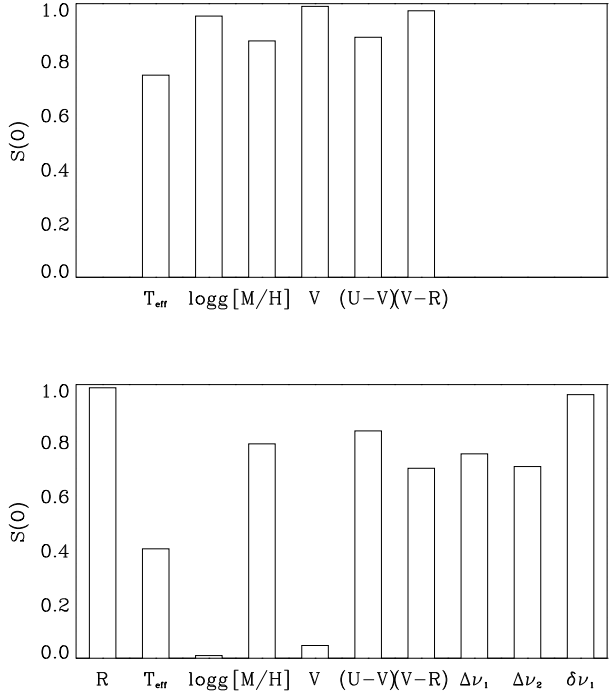}
\caption{{\sl Top:} The significance of the observables without
oscillation information and radius measurements 
using $\epsilon(R)$ = 0.01 R$_{\odot}$ and $\epsilon(\nu)$ = 1.3$\mu$Hz.
{\sl Bottom:} The same but including oscillation and radius
measurements.
\label{fig-signif_obs}}
\end{figure}

Figure~\ref{fig-signif_obs} (top panel) illustrates the importance of some 
typical observables in the absence of a radius measurement 
and oscillation frequencies.
The $\epsilon_i$ are those quoted from Table~\ref{tbl-observables}.
While this figure is mainly for comparison with 
the lower panel we highlight a few points:
the photometric observables appear to supply more information
than the spectroscopic --- this is contrary to what is often believed.
For example,
the colors are more significant than $T_{\rm eff}$. 
S$([$M/H$])\sim 0.85$ where we may have assumed that most 
information about chemical composition comes from $[$M/H$]$
and so should be $\sim 1$.
Also note the high values of both S($\log g$) and S($V$). 
They are responsible for 
determining both the radius and the mass of the star.

We then included $R_{\star}$, {\it two} $\Delta \nu_{\rm n,l}$ and {\it one} 
$\delta \nu_{\rm n,0}$ in the previous
set of observables.
Figure~\ref{fig-signif_obs} (lower panel) illustrates how the 
relative information
content in each $O_i$ changes.
There are two notable changes:
(1) $\log g$ and $V$ contain almost no information while both
the observed $R_{\star}$ and $\Delta \nu_{\rm n,l}$ 
become responsible for determining the true radius
(and mass) of the star, (2) 
S$(V-R)$ and S($[$M/H$]$) decrease by a small amount indicating that
the information from colors and metallicity
is not contained in the new observables.
We believe this information is primarily the 
chemical composition.
This implies that either colors or a metallicity
are important observables to have when 
complementing  seismic and radius measurements.
We also note that while S$(V-R)$ decreases 
by $> 25\%$, S$(U-V)$ remains the same.

\begin{figure}
\center{\includegraphics[width = 9cm]{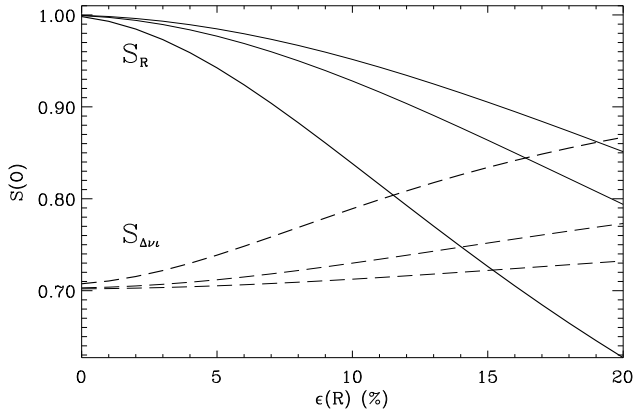}}
\caption{The change in the S($R_{\star}$) and 
S($\Delta \nu_{\it n,l}$) as we vary $\epsilon(R)$.
$\Delta \nu_{\it n,l}$ increases in significance as $R_{\star}$ 
becomes more poorly constrained.
Bold lines indicate $\epsilon(\nu)$ = 0.5$\mu$Hz, lighter
lines indicate $\epsilon(\nu)$ = 1.3, 2.5 $\mu$Hz.
\label{fig-signif_dlv}}
\end{figure}

We included more than one $\Delta \nu_{\rm n,l}$ to 
highlight how when we change the measurement errors, 
the information contained
in one observable is passed to another,
in particular, between $R_{\star}$ and $\Delta \nu_{\rm n,l}$.
If we had included just one $\Delta \nu_{\rm n,l}$,  then
S($\Delta \nu_{\rm n,l}$)~$\sim$~S($\delta \nu_{\rm n,0}$)~$\sim$1.
When there are two observables contributing
similar (but not the same) information, 
the content gets spread among them.
Similarly, with three $\Delta \nu_{\rm n,l}$, 
S($\Delta \nu_{\rm n,l}$) $\sim$ 0.55.
Figure~\ref{fig-signif_dlv} illustrates this.
S($\Delta \nu_{\rm n,l}$) (bold dashed line) and 
S($R_{\star}$) (bold continuous lines)
are shown as 
a function of $\epsilon(R)$ with $\epsilon(\nu) = 0.5\mu$Hz.
At the smallest $\epsilon(R)$, S($R_{\star}$)$\sim$1.
Increasing $\epsilon(R)$ causes S($R_{\star}$) to decrease, 
while S($\Delta \nu_{\rm n,l}$) 
increases because it is taking over its role.
The non-bold lines show the same but for 
$\epsilon(\nu) = 1.3,2.5\mu$Hz.
There is a similar but slower trend, because the radius remains
more important than $\Delta \nu_{\rm n,l}$ even for higher
values of $\epsilon(R)$.
Using only one $\Delta \nu_{\rm n,l}$ would not show an increase
in S($\Delta \nu_{\rm n,l}$) because it would be already near 1,
but S($R_{\star}$) would decrease more slowly, while
observables such as $\log g$ and $V$ would increase in importance.

Both Figures~\ref{fig-signif_obs} and \ref{fig-signif_dlv} show that
the relative importance of observables depends not only on the
individual $\epsilon_i$ but also on the combination
of $O_i$ available, and that there is no straightforward
relationship between a set of $O_i$ and the
parameters they are responsible for constraining.

\subsection{Parameter Correlations \label{sec-parametercorrelations}}

The combination of observables we use is important
for the independence of the parameters.
In general, there should be a unique set of
frequencies for every possible combination
of $(M,\tau,X,Z,\alpha)$.
However, we are not using individual
$\nu_{\rm n,l}$ but frequency separations for the reasons
explained in Section~\ref{sec-observables}.
This creates some dependencies among the parameters
such as that between $M$ and $\tau$, $M$ and $\alpha$
and $M$ and $X$.
With the addition of measurement errors, this
degeneracy between different parameter combinations
becomes worse.

To illustrate some of the dependencies among the 
parameters, we calculated $\chi^2$ surfaces for
two parameters $p$ and $q$ while keeping the other parameters
fixed at their correct values.
We evaluate 68.3\%, 90\% and 99\% confidence levels
as defined for a two-dimensional $\chi^2$
i.e. 
$\Delta \chi^2_r = \chi^2_{min} - \chi^2_{pq} = [2.31,4.61,9.21] $.
We used $\epsilon(R)$ = 0.01 R$_{\odot}$ and
$\epsilon(\nu)$ = 1.3$\mu$Hz and defined
$\chi^2_r = \chi^2/(\#observables -2)$.

Using a set of $\nu_{\rm n,l}$ as observables the 99\%
confidence region encircled the correct parameter
values to within less than 1\% of each of the 
corresponding parameters (it produced almost a dot 
centered on its true value),
supporting the idea
that each set of parameters should produce
a unique set of $\nu_{\rm n,l}$.

\begin{figure}
\center{\includegraphics[width = 9cm]{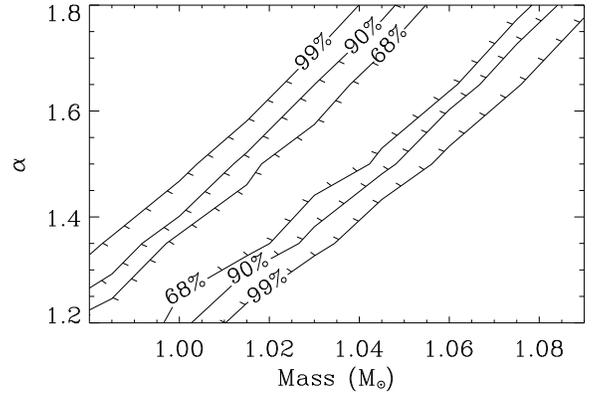}}
\caption{$\chi^2$ contour plot for $M$ and $\alpha$.
$(X,Z,\tau)$ are fixed at their correct value.
We used OS3 (see Section~\ref{sec-properrors}) 
with $\epsilon(R)$=0.02 R$_{\odot}$ 
and $\epsilon(\nu)$ = 1.3$\mu$Hz to calculate $\chi^2_r$.
\label{fig-chi2_ma_30}}
\end{figure}

To show an example of the dependencies introduced by using 
the frequency separations, 
Figure~\ref{fig-chi2_ma_30}
illustrates the correlated $\chi^2$ surface for mass 
and $\alpha$ while holding ($\tau$, $X$, $Z$) fixed
at their true values using OS3 (see section~\ref{sec-properrors}).
Clearly for any value of $M$  there is a corresponding $\alpha$ that
will give a minimum $\chi^2$ value.
We hope that a minimization will lead to accurate
values, but we may need additional observables
to help constrain the solution.

\subsection{Propagated errors \label{sec-properrors}}

Can we get a precise determination of the mass using a radius measurement?
How well do we need to know our observables?
Will the effort of obtaining an interferometric
diameter be outweighed by the benefits,
or can some other observable produce a similar result?

To answer these questions, we investigated the expected parameter
uncertainties through equation (\ref{eqn-covariance}) 
for various observables with different errors.
For the remainder of this paper we shall discuss three sets of 
observables: 

$${\rm OS1}=\{R_{\star}, T_{\rm eff}, [{\rm M/H}], \bar{\Delta \nu}, \bar{\delta \nu}\}, $$
$${\rm OS2}=\{R_{\star}, T_{\rm eff}, [{\rm M/H}],\Delta \nu_{\rm n,l} \} $$ 
and
$${\rm OS3}=\{R_{\star}, T_{\rm eff}, [{\rm M/H}],\Delta \nu_{\rm n,l}, \delta \nu_{\rm n,l} \} $$ 
for $n=9,10,...27$ and $l=0,1,2$.
For OS2 and OS3 we include a total of 21 large frequency separations, 
where we have assumed
that we cannot identify every consecutive mode for each $l$ and
thus have even fewer frequency separations.
For OS3 we also include 5 small frequency separations.
We remind readers that the parameter uncertainty is denoted by
$\sigma$ while the measurement errors are denoted by $\epsilon$.

For ease of reading we shall refer to $\epsilon(\nu) = 0.5 \mu$Hz as LN
and 1.3$\mu$Hz as HN.
Figure~\ref{fig-theoretical_errors} shows the theoretical uncertainties 
on each of the parameters as a function of $\epsilon(R)$.
We show results using LN (black) and HN (gray).
The dashed lines show the results for OS1.
The continuous lines show OS2, and the dotted lines show OS3.

For all parameters we find that $\sigma(P)$ (parameter 
uncertainty)
is a function of $\epsilon(R)$ for all combinations of observables.
We also see that for $M$, $X$ and $Z$ the different observable
combinations and errors do not impact significantly the
theoretical uncertainties for $\epsilon(R) \le 3\%$.
Only for $\tau$ and $\alpha$, using OS2 instead of OS1 may not
always be beneficial.  
This is because $\delta\nu$ is excluded from OS2 and this observable 
is essential for determining $\tau$ and, as it turns out, $\alpha$ too.

\begin{itemize}
\item {\sl Mass}: 
Whatever combination we choose, we find that for $\epsilon(R) \le 3\%$
$\sigma(M)$ does not change.  
A radius measurement will always be more important than 
seismic information once we reach a precision of this order.
As the radius error begins to increase, its relative weight
for mass determination decreases, while the relative weight of the seismic
information increases.  
The quantitative differences between
using LN and HN
for each observable combination  for $\epsilon(R) \ge 3\%$ 
supports the idea that seismic information is 
the dominant contributor to the reduction of the 
uncertainty in mass at higher $\epsilon(R)$.

\item {\sl Age}:
Either a reduction in frequency error or an increase in observables
significantly affects $\sigma(\tau)$.
OS1 and OS3 do not benefit significantly from the radius measurement,
but both produce better constraints on the age than OS2.  This is 
precisely  because they
contain $\delta\nu$ which is fundamental in
constraining this parameter.
However, for OS2 a precise radius measurement reduces $\sigma(\tau)$ by 200\%
for LN and HN. The lack of the observable $\delta \nu$ 
forces $R_{\star}$ to constrain the age.
Not surprisingly,  we find that if we use LN instead of HN for all
observable combinations
there is a large improvement in $\sigma(\tau)$.

\item {\sl X}:
Adding observables and obtaining smaller frequency errors
leads to subtle differences in $\sigma(X)$ for low radius
error.
OS1 with either LN or HN produce very similar constraints on $X$.
Using OS2 does not change $\sigma(X)$ very much for $\epsilon(R) \le 3\%$.
At larger values, we find that the additional large frequency separations 
(especially for LN) constrain $X$ to about 10\%, an improvement of a factor
of 2 over HN.
OS3+HN and OS2+LN produce similar results, indicating that it is not the
presence of $\delta\nu$ specifically that is important, just the addition 
of (seismic) observables.

\item {\sl Z}:
This is one of the parameters that is least influenced by either radius or
seismic observables.  It is mainly constrained by $[$M/H$]$, but because 
this implies knowing $X$, other observables play a role in
constraining this parameter.
Radius is most important for $Z$ when we use OS1, because
we have less information available.
Using OS2 and OS3 decreases the dependence of $Z$ on $R_{\star}$. 
In fact, with OS3+LN, we find that the radius has almost no 
role in constraining $Z$.
For $Z$, the radius is important only when there is
not enough information available.  As we increase the number
of seismic observables, this information
supercedes the role of the radius.  We also find that 
whatever combination we use, $\sigma(Z)$ never decreases
below 10\% even using LN.

\item {$\alpha$}:
The determination of $\alpha$
seems to be more complex than the
determination of any other parameter.  In general, the addition 
of observables (in number) coupled with better frequency errors
leads to lower parameter uncertainties. This is not always the 
case for $\alpha$.
OS2 is least influenced by a reduction in radius error.
In fact, we find for HN, OS1 gives better constraints on $\alpha$
than OS2 for $\epsilon(R) \le 4\%$.
Only when an observable set contains the small frequency separation
and for small values of radius error, do we get a significant 
decrease in the theoretical uncertainty in $\alpha$.  
Using LN instead of HN always affects the determination
of $\sigma(\alpha)$, but more so for OS2 when the radius information
is less important than the frequency information.

\end{itemize}

The role of the radius sometimes depends on what other information
is available.  For example, if we seem to be {\it missing} 
information, then improving the radius error will lead to a corresponding
reduction in the uncertainty of some parameter, like $Z$.
The determination of $\alpha$ shows an interesting trend. A precise radius
determination {\it complements} precise frequencies.  Either of these
alone does improve the uncertainty, but both together have a larger impact 
on $\sigma(\alpha)$.
Clearly for $\epsilon(R) < 3\%$ the choice of observable set 
does not change $\sigma(M)$; it is $R_{\star}$ which dominates 
the mass determination.

\subsection{Magnitude versus Radius}

Are these uncertainty values worth the observing time? 
We compared the propagated error using 
OS2 with $M_V$ (magnitude) instead of $R_{\star}$ (radius) 
to see whether $M_V$ could 
produce results similar to $R_{\star}$.
$M_V$ is the only other measurement
that
might constrain the mass as $R_{\star}$ does.

Figure~\ref{fig-mag_v_mass}
shows $\sigma(M)$ as a function of both
$\epsilon(R)$ (continuous lines) and
$\epsilon(M_V)$ (dashed lines) 
for two $\epsilon(T_{\rm eff}) = [50,230K]$ 
for LN (left panel) and 
HN (right panel).
We have attempted to scale the expected errors  on both
observables so that the quoted errors are comparable 
in terms of measurement difficulty.
We chose $\epsilon(R)$ = 0.01 R$_{\odot}$ ($\sim 1\%$) to be similar
to $\epsilon(M_V) = 0.05$ mag.
For both cases $\sigma(M)$ using $M_V$ will never reach the
precision that a radius measurement allows.
The smallest mass uncertainty using $M_V$ is $\sim 5\%$, while
if we use $R_{\star}$ it reaches below 1\%.
Also $M_V$ needs to be complemented with other good observables
such as $T_{\rm eff}$ and LN 
to get this precision, whereas
$R_{\star}$ is a completely independent measurement that 
does not need the help of other observables.
These are two very clear reasons for choosing to obtain
an interferometric measurement of diameter.
For HN (right panel)
it is even clearer that $R_{\star}$ is most important.
 
Even if our scaling is incorrect for the
comparison of magnitude and radius errors,
it is unlikely that an $M_V$ measurement will 
have a precision of better than 0.05 mag, in which
case the smallest mass uncertainty is really
above 7\%.

\section{Simulations \label{sec-simulations}}

To test the validity of the theoretical results presented in 
Section~\ref{sec-properrors} we performed simulations
and tried to fit the observables to recover the
input parameters.
Part of this work also involved finding a reliable method to apply
to any set of observables
so that we could succesfully estimate the parameters
of the star within the theoretical uncertainties.
We use two methods: direct inversions and minimizations.
The direct inversions provide a test for the linear 
approximation of equation~(\ref{eqn-taylor}) by using an initial
guess of the parameters {\bf x$_0$} and
equation~(\ref{eqn-solution}).  We did this only for OS1.
Minimizations
iteratively incorporate a modified version 
of equation~\ref{eqn-solution} (Levenberg-Marquardt algorithm)
and were performed with OS1, OS2 and OS3.



The observables 
${\bf O} = \{R_{\rm o},T_{\rm o},[{\rm M/H}]_{\rm o},\nu_{\rm (n,l)o}\}$
were simulated as follows:
\begin{equation} 
R_{\rm o} = R_{\rm r} + r_{\rm g} \epsilon(R)
\label{eqn-simrad}
\end{equation}
\begin{equation} 
T_{\rm o} = T_{\rm r} + r_{\rm g} \epsilon(T)
\label{eqn-simteff}
\end{equation}
\begin{equation} 
{\rm [M/H]}_{\rm o}  = [{\rm M/H}]_{\rm r} + r_{\rm g} \epsilon([{\rm M/H}])
\label{eqn-simmh}
\end{equation}
\begin{equation} 
\nu_{i\rm {o}} = \nu_{i\rm {r}} + r_{\rm g} 
\epsilon(\nu) +\left ( \frac{\nu_i}{\nu_0} \right)^2
\label{eqn-simnu}
\end{equation}
where $R = R_{\star}$; $T = T_{\rm eff}$;
the subscipt 'o' denotes the simulated {\it observed} value;
the subscript 'r' denotes the real value that comes directly from the
stellar code; the $\nu_i$ are individual frequencies;
$r_{\rm g}$ is a random
Gaussian error for each observable with $\sigma = 1$;
$\nu_0$ is an arbitrary reference frequency value, which was
chosen to be 6000$\mu$Hz.
Using these simulated frequencies we calculated each 
of the $\Delta \nu_{\rm n,l}$ and $\delta \nu_{\rm n,0}$ from 
equations (\ref{eqn-largeseparation}) and (\ref{eqn-smallseparation}).
Table \ref{tbl-observables} provides the errors, and 
$\epsilon(\Delta\nu_{\rm n,l}) = 
\epsilon(\delta\nu_{\rm n,0}) = \sqrt{2\epsilon(\nu_{\rm n,l})^2}$.


\subsection{Linear Approximation -vs- Minimizations for OS1 \label{sec-paresults}}

Direct inversions will allow us to investigate the natural
weight that $R_{\star}$ and its error have on the 
determination of $M$, because we do not use
a minimization algorithm but just the linear approximation and 
equation~(\ref{eqn-solution}).
For each radius and frequency error, we generated a total of 100 simulations
of the observables.  We obtained a 
list of ${\bf x_0}$ from 
a grid of models (Section~\ref{sec-modelgrids}) and 
chose the first set from this list to be the initial guess.
For most of the simulations, an initial mass of 
$\sim 0.9 M_{\odot}$ was used (the real value = 1.03).
The system is relatively linear in these parameters and 
observables hence the inversion results should be 
independent of the initial values.
(This was tested, and the results varied to less than 1\%.)
The simulations were repeated for various frequency
and radius errors.

Figure~\ref{fig-di_simulationresults} shows the 
mean and standard 
deviation of the inverted ({\it fit}) mass from the simulations as a 
function of the radius error.  
The dotted line shows the real mass value.
Each panel corresponds to a different $\epsilon(\nu)$.
The upper panel shows LN, the center
HN and the lower panel has $\epsilon(\nu)$ = 2.5$\mu$Hz.
The envelope dashed lines show the theoretical uncertainties.

Figure~\ref{fig-di_simulationresults} (top and center panels) shows
that the mass uncertainty is a clear linear function of
the error in radius.
In fact, 
$\epsilon(\nu)$ has a very modest effect on both the precision and
the accuracy of the mass.
For the two upper panels, there is no 
apparent difference, indicating that
the radius primarily determines
the mass.
However at $\epsilon(\nu)$ = 2.5$\mu$Hz we do see
an effect on both the precision {\it and}
the accuracy of the results.
This appears to be a side effect of poorly constraining
the other parameters, which indirectly leads to incorrect mass
estimates.

It is encouraging to find that
the simulation results are consistent with theory {\it and} that
we can safely estimate the mass to within its theoretical 
uncertainty.
Unfortunately for $\epsilon(R) = 5\%$, {\it safely} estimating the
mass to within 20\% may not be very useful.  However, 
we do find that for $\epsilon(R) \le 2\%$ we
are capable of determining the mass better than the theoretical
uncertainty and quite accurately (for the two smaller frequency errors).
For all simulation results, the true mass value always falls within
1 standard deviation of the fit mass.

For the other parameters, the theoretical predictions
are consistent with the simulation results.
Figure~\ref{fig-di_age_3nus} shows the mean and the 
standard deviation of the inverted ({\it fit}) age.
As expected, the size of the frequency error is very important, 
especially for OS1.
$\epsilon(R)$ does have a modest effect and is most 
noticable for LN (top panel).

To test whether a minimization algorithm 
improves the 
uncertainties on the parameters, we
repeated the simulations and then fit for the parameters (using the minimizing
method)
for a range of radius errors and LN. 
Figure~\ref{fig-simul_errors_obs1} shows
the standard deviations of the fit parameters, 
(for age we show $\sigma(\tau)/4$).  All of the
parameter's real values were within the 
$\sigma$ shown in this figure.

The precision of the parameters are better than
the theoretical ones 
(Figures~\ref{fig-theoretical_errors},\ref{fig-simul_errors_obs1}).
There is a clear dependence of the uncertainties on the radius
error when $\epsilon(R)$ $<$ 3\%.
Unlike the theoretical uncertainties, 
for larger $\epsilon(R)$ there is no significant degradation in
the precision of the parameters.
This uncertainty-error dependence changes drastically for
larger $\epsilon(R)$ because the minimization method is forcing
a balance between the errors and the
observations, unlike when we use the simple linear approximation
(Equation~\ref{eqn-solution}).

\subsection{Simulations and Minimizations with OS2 and OS3 \label{sec-simulations2}}

Since we are introducing more frequency observables in OS2 and OS3, 
we are effectively giving less weight to the radius
measurement for constraining the parameters.
The fit parameters are thus more sensitive to the
initial parameter estimates.
It is therefore necessary to use small
grids of models to obtain a good initial guess,
and then let the minimization algorithm fit all of the
observables.

\subsubsection{Model Grids \label{sec-modelgrids}}

We created grids of models that span the five dimensional model
space over a range of values that encompass about 10-30\% on 
either side of the parameter values given in 
Table~\ref{tbl-parameters}.
These values covered $0.95 \leq M \leq 1.09$ M$_{\odot}$,
$0.1 \leq \tau \leq 2.0$ Gyr,
$0.68 \leq X \leq 0.76$,
$0.014 \leq Z \leq 0.022$ and
$1.2 \leq \alpha \leq 1.8$.
We divided each of the parameter ranges into five 
(ten for mass), and calculated models for
each of these points, and then interpolated only
along the mass for a much
finer grid. 
We are restricting the range in mass, age and $Z$, but
we assume that we can constrain the parameters
within this range from the observables.

We later extended the mass range to be between 0.75 and 1.20 M$_{\odot}$ to
investigate whether our original parameter range was too restricted.
Using  $\tau=0.1,X=0.70,Z=0.024, \alpha=1.5$
and varying mass, we used the grids to see how well we
could recover the mass in this extended mass range.
These parameters were chosen to test for 
any biases introduced by selecting our original
ranges to be nearly symmetric about the correct values.
We tested this for $\epsilon(R)$ = 0.02 R$_{\odot}$ and HN only.
There were 100 simulations for each mass.

We searched through the grids to find all of the parameter combinations
that reproduced $R_{\star}$ and $\bar{\Delta \nu}$ to within $\pm 3 \epsilon$.
We discarded the combinations with $\chi^2$ larger than 1, and
adopted the mean value of each parameter from the remaining combinations 
to be the {\it fit} values.  We later used them as the initial
parameter values for the minimization.
Figure~\ref{fig-grid_mass} shows the 
mean
of the simulations $\pm$
$3\sigma$ where $\sigma$ is the standard deviation of each
group of simulations.
The dotted line indicates the
ideal position for the mean of the simulation results.
While there seems to be some systematic offset as a function
of mass, in most cases the $3\sigma$ crosses the 
correct value.
For the other parameters, the fit values are
not as well determined (we haven't interpolated for them).
This appears to be responsible for the offset in fit mass.

\subsubsection{Results of Minimizations}

Initial estimates of the parameters were obtained from the model grids.
We added a constant of 0.02 to the initial estimate of $X$ for two reasons:
1) $X$ had always shown a systematic offset when using the model grids 
and, 2) if the value of $\chi^2$ from our initial
estimates is too low, the minimization algorithm may not search among all
possible solutions.
We then allowed the code to minimize in all available
observables.
The solution usually converged within four iterations.
We set the $\chi^2_r$ tolerance level to be 0.1,
where $\chi^2$ is calculated from 
equation~(\ref{eqn-chi2}), but we used
3$\epsilon$ errors for the process instead of 1$\epsilon$,
and $\chi^2_r = \chi^2 / n$ where $n$ is the number
of degrees of freedom.
An ideal value of $\chi^2_r$ is 1. 
Anything much larger implies a bad fit,
an error in the physics, or incorrectly
quoted errors.

$\epsilon(R)$ took values from 0.001-0.05R$_{\odot}$ ($\sim 0.1-6.0\%$).
$\epsilon(\nu)$ took values of 0.5 (LN) and 1.3$\mu$Hz (HN).
Figure~\ref{fig-simulations_all} shows the mean and 
standard deviation (1$\sigma$) of the fit mass 
as a function of $\epsilon(R)$ for OS2+HN.
The dotted line is the true mass value.
We show only this combination because those for 
LN\footnote{Extending the radius error to 0.001 R$_{\odot}$
did produce a small improvement in the parameters at this level for LN.}
and those using OS3 did not lead to a significant
difference, 
except for $\tau$ and $\alpha$ whose uncertainties are
represented by the dashed lines in Figure~\ref{fig-sim_alpha_tau}. 

Our global results show that for all of the parameters
1) the mean values fit to within 1$\sigma$
of the correct value (Figure~\ref{fig-simulations_all} shows this for
$M$ only).
2) The parameter uncertainties are 
smaller than those predicted by theory.  This is highlighted in
Figure~\ref{fig-sim_alpha_tau}.
3) All of the uncertainties show some dependence on $\epsilon(R)$.

\begin{figure}
\center {\includegraphics[width = 9cm]{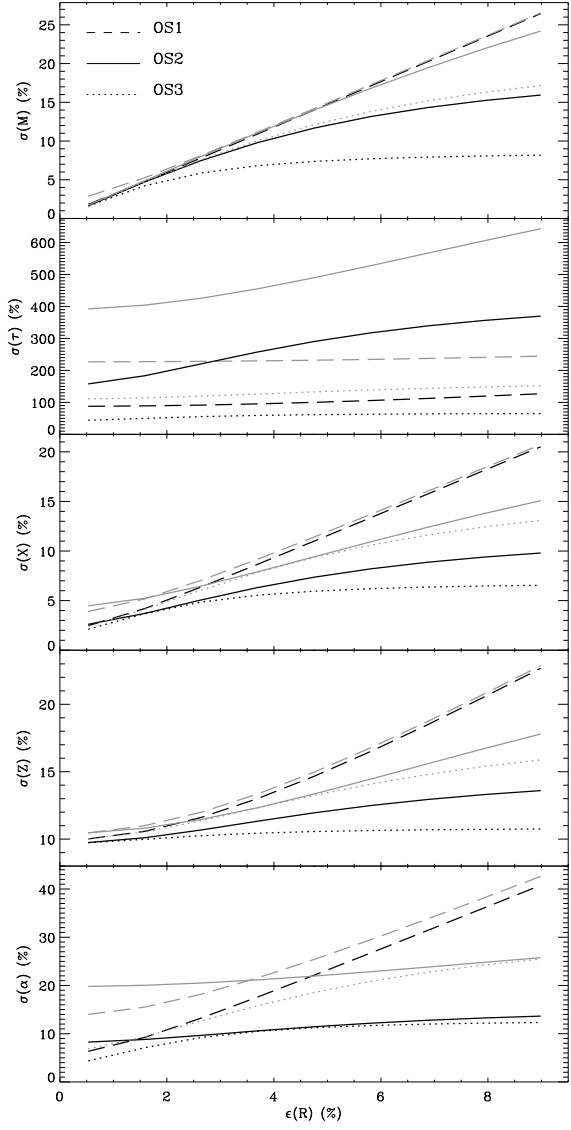}}
\caption{Comparison of the expected theoretical uncertainties 
between OS1 (dashed), OS2 (continuous) and OS3 (dotted)
observable sets and different frequency errors 
for all parameters as a function of $\epsilon(R)$.
Results for HN are shown by the gray line, and those
for LN are represented by the black lines 
(darker lines = smaller frequency error).
\label{fig-theoretical_errors}}
\end{figure}

\begin{figure*}
\center{\includegraphics[width = 18cm]{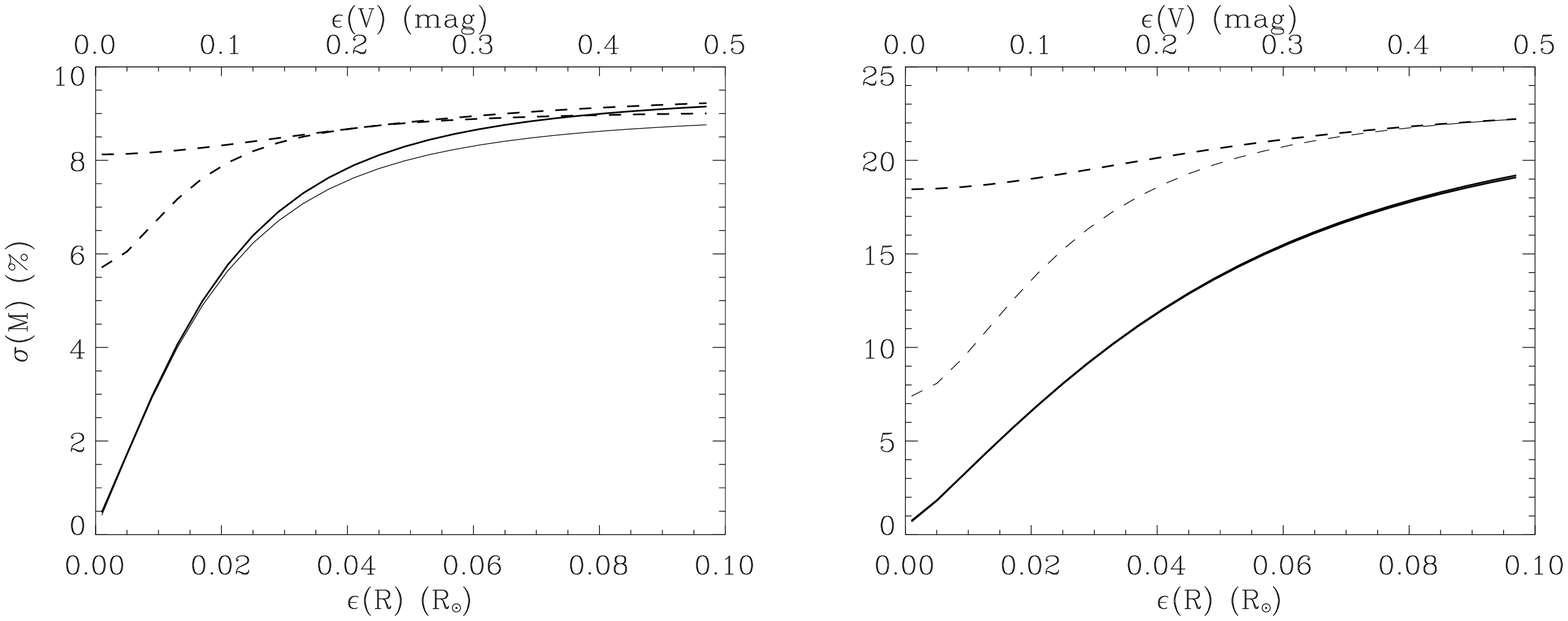}}
\caption{Comparison of propagated mass error using either
$R_{\star}$ (continuous) or $M_V$ (dashed) with OS2 observables at 
two $\epsilon (T_{\rm eff}) = [50,230K]$.
The left panel is that when using $\epsilon(\nu)$ = 0.5$\mu$Hz,
the right is that for $\epsilon(\nu)$ = 1.3$\mu$Hz.
The lower x-axis shows $\epsilon (R)$ in appropriate units whereas
the upper x-axis shows what we could consider a comparative
$\epsilon (M_V)$. \label{fig-mag_v_mass}}
\end{figure*}

Mass is the parameter that is most 
affected by an improvement in the precision of $R_{\star}$.
Figure~\ref{fig-simulations_all} shows the mean and standard deviation
of the fit mass. 
$\sigma(M)$ is clearly dependent on $\epsilon(R)$ for $< 3\%$ errors.
If we reach 0.1\% in radius error, then we can determine the mass to
better than 1\%.
For $\epsilon(R) > 3\%$, $\sigma(M)$ does not change significantly.  
This can be seen more clearly in Figure~\ref{fig-sim_alpha_tau}.
For  $\epsilon(R)$$ < 3\%$, there is little difference between 
the uncertainties arising from the minimizations (and more observables) 
or the linear approximation with 5 observables 
(Figures~\ref{fig-di_simulationresults},\ref{fig-simul_errors_obs1}).
This can be understood by considering that for small
radius errors, this observable has such a large 
weight that it determines uniquely the 
mass value.  
In this regime, the different combinations of observables
 do not have any influence,
just as the theoretical uncertainties  showed
(Figure~\ref{fig-theoretical_errors} top panel).
So whether we use a minimization method with 26 observables or a 
direct inversion with only 5 observables, having a well
measured radius will uniquely determine the mass.  
When the relative weight of the radius decreases, i.e.
its error increases, 
other observables begin to play a role.  
This is why we see a dramatic change in the mass uncertainty 
between using OS1, OS2 and OS3 when 
$\epsilon(R)$ $> 3\%$
(Figures~\ref{fig-di_simulationresults},\ref{fig-simulations_all}).

Figure~\ref{fig-sim_alpha_tau} shows the standard deviations for 
all of the parameters.  The dashed lines represent the results for OS3+LN and
the dotted line for OS3+HN 
(continuous are OS2+HN).  For $M$, $X$, and $Z$ the results
did not change significantly.
\begin{itemize}
\item 
The top panel illustrates $\sigma(\tau)$.  The values are 
rather high, but the standard deviations that result from a star
of 4 Gyr (instead of 1 Gyr) have similar 
absolute values and thus much smaller relative values. 
Note the qualitative differences
between the set of observables with $\delta \nu$ (dashed and dotted lines)
and the set without (continuous line).  
Even if we use 21 large frequency separations with the
precision of LN, we do not get the same age uncertainty as when
we use OS3.
This demonstrates that $\delta \nu$ is crucial for obtaining
a precise determination of the age (or stellar evolutionary stage)
(e.g. \citet{bro94a,bro94b,mig05,maz05}).
For $\epsilon(R)$ $<2\%$ we also see a slight drop in $\sigma(\tau)$.
This is because for small radius errors, we get precise determinations
of both $M$ and $X$, so the range of possible ages that fit 
$\delta \nu$ is restricted, and hence we get better age determinations.

\item The lower panel shows the standard deviations of all of the 
other parameters.  Both $Z$ and $\alpha$ exhibit a small 
$\epsilon(R)$ dependence: 9-12\% for $Z$ and 7-10\% for $\alpha$
over the radius error range.
Theory (Figure~\ref{fig-theoretical_errors}) predicted 
that $\sigma(Z)$ would never fall below 10\%, and
increase only slightly to 12-14\% at larger radius errors.
This trend is clearly consistent with the simulation results.
$\sigma(\alpha)$ was predicted to be 20-22\% over the range of radius
error for OS2+HN and 8-14\% for OS2+LN.  The simulations
show no significant differences between these observable combinations.
For OS3+LN, theory predicted 4-12\% uncertainties, comparable to the
dashed line in Figure~\ref{fig-sim_alpha_tau}.
$\sigma(\alpha)$ benefits significantly from the additional 
$\delta \nu$ reaching a precision of about 4\% and slowly degrading 
with $\epsilon(R)$ until $\epsilon(R)$ = 3\% when it levels off at a  
$\sigma$ value comparable 
to the other observable combinations.
$\sigma(X)$ shows a dependence on radius error similar to $\sigma(M)$.
For $\epsilon(R)$ $\le 3\%$ we generally estimate $X$ to within 2-3\%.
For larger radius errors, $\sigma(X)$ does not increase, indicating
the redundancy of a radius measurement after $\epsilon(R)$ = 3\%.

\end{itemize}

While some parameters may not seem very dependent on $\epsilon(R)$,
having a radius observable allows us to 
estimate some parameters  correctly. This in turn
leads to a better estimation of the other {\it non-radius dependent} 
parameters because
of the more restricted acceptable range.

These results stem from an automatic method of 
fitting the observations. Given that there is no
human input, we are confident in the
results and expect that even more precise
results will be possible by paying attention to each individual case, and
individual observable $\chi^2$ values.

\section{Conclusions \label{sec-conclusions}}

We investigated the role that a precisely measured
radius, such as that obtained through interferometry,
has on the determination of the mass of a star.
We also looked at how radius and oscillation frequencies
work together to determine other stellar parameters.

\begin{figure}
\includegraphics[width = 9cm]{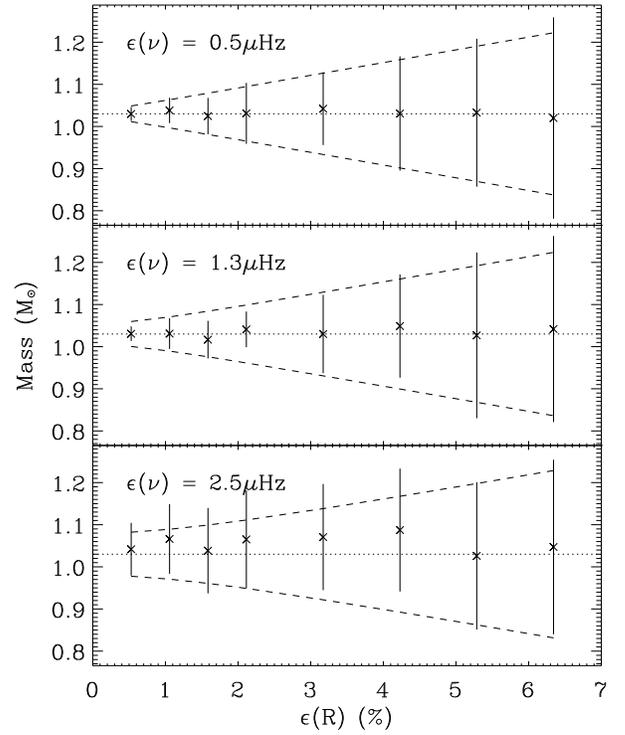}
\caption{Mean and standard deviation of the fit ({\it inverted}) mass
for the simulations as a function of $\epsilon(R)$ using OS1.
{\sl Upper} Results for LN;
{\sl Center} Results for HN;
{\sl Lower} Results for $\epsilon(\nu)$ = 2.5$\mu$Hz.
\label{fig-di_simulationresults}}
\end{figure}

\begin{figure}
\includegraphics[width = 9cm]{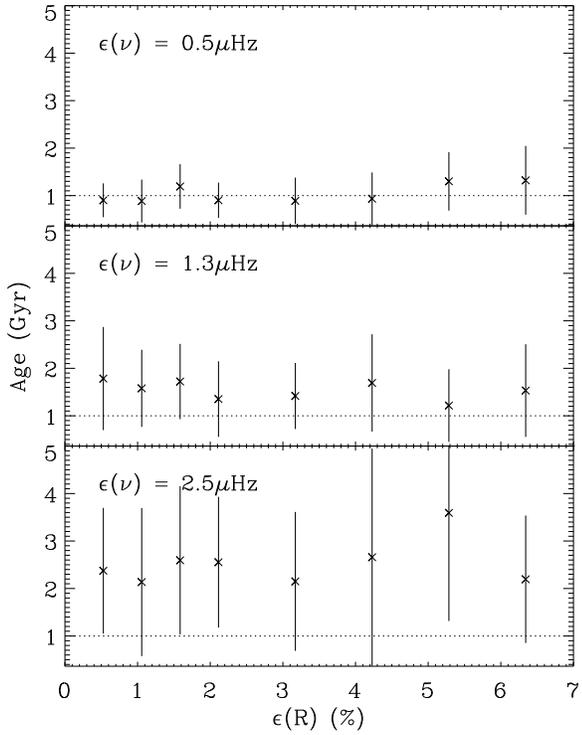}
\caption{Similar to Figure~\ref{fig-di_simulationresults}, but 
for age.
\label{fig-di_age_3nus}}
\end{figure}

\begin{figure}
\includegraphics[width = 9cm]{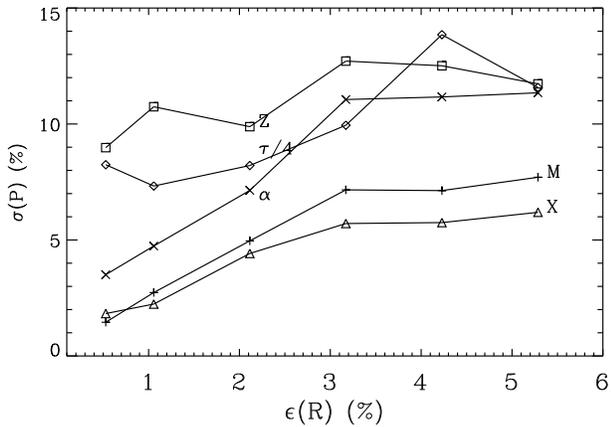}
\caption{Parameter uncertainties as a function of 
$\epsilon(R)$ determined from the minimizations 
using OS1+LN.}
\label{fig-simul_errors_obs1}
\end{figure}

\begin{figure}
\includegraphics[width = 9cm]{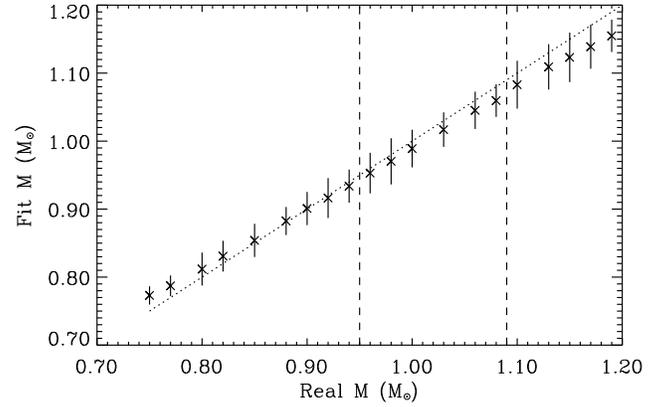}
\caption{Mean and 3$\sigma$ uncertainty of the
fit mass as a function of the true mass value
resulting from the simulations using only the
model grids.  
Dashed vertical lines indicate boundary of extrapolated and 
evaluated models.
\label{fig-grid_mass}}
\end{figure}

\begin{figure}
\includegraphics[width = 9cm]{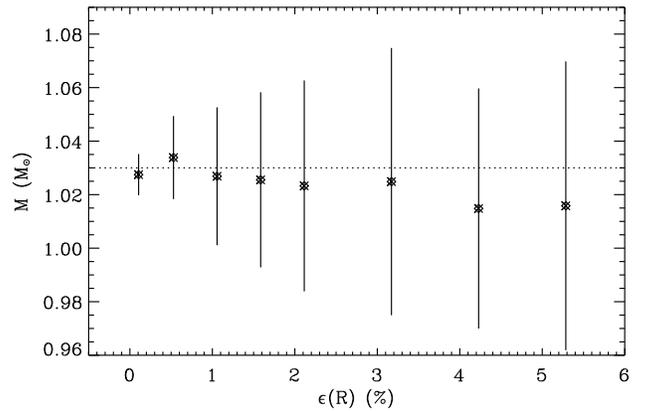}
\caption{
Mean and 1$\sigma$ standard deviation of the minimizations 
for the fit mass using OS2+HN as a function of $\epsilon(R)$.
\label{fig-simulations_all}}
\end{figure}

\begin{figure}
\includegraphics[width = 9cm]{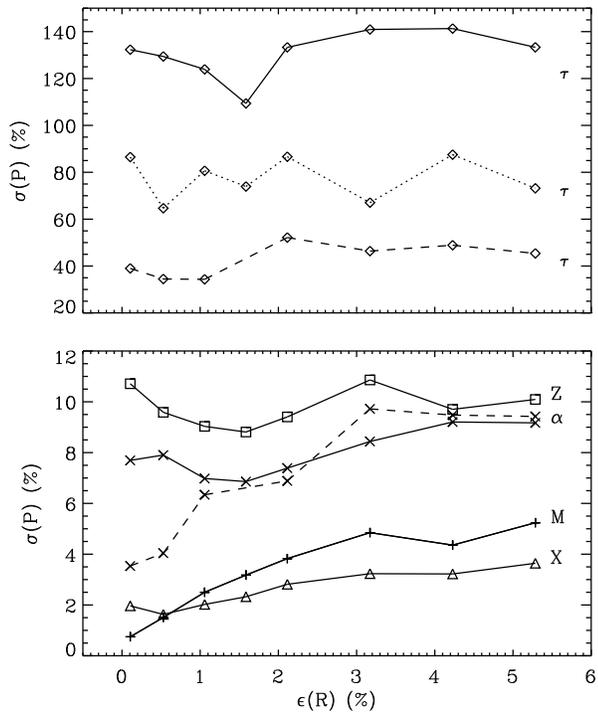}
\caption{Comparison of standard deviations of all of the fit 
parameters as a function of $\epsilon(R)$.
OS2+HN, OS3+LN and OS3+HN ($\tau$ only) are denoted by the solid, dashed and
dotted lines respectively.
\label{fig-sim_alpha_tau}}
\end{figure}

\begin{itemize}

\item
We found that the importance of a radius measurement 
depends on the combination of 
the available observables {\it and} their corresponding errors.
For some typical observables such as $R_{\star}$ and $\Delta \nu_{\rm n,l}$, we
can expect a mass uncertainty of between 1 and 4\%
for $\epsilon(R)$ between 0.1 and 3\%
(keeping in mind that these numbers stem from an
automatic parameter search and in a real case study we would
pay more attention to each observable and can expect better precision.)
For $\epsilon(R)$ $\le 3\%$ the mass is uniquely determined by the radius
observable, allowing us to use the frequencies just
to probe the stellar interior.

\item
Our  simulation errors are consistent with the theoretical errors when we use
the linear approximation (Figure \ref{fig-di_simulationresults}).  
Using a minimization
method yields smaller than predicted uncertainties with similar
qualitative results for all values
of radius error, but particularly when $\epsilon(R) > 3\%$.
For these values, 
the weight of the radius measurement decreases, allowing
other observables to play a greater role in the determination 
of the parameters.  The minimization method allows 
an optimal trade-off between all observables and 
their corresponding errors.

\item
We  emphasize the importance of understanding how each observable 
contributes to the determination of each parameter.  
For example, if we want to get an estimate of $Z$ better than 10\% we need 
a more precise $[$M/H$]$.
A radius measurement would not be useful. 
To correctly determine age (or stellar evolution stage) it is crucial
to observe $\delta \nu$.  Even with 21 large frequency spacings, we will not
determine the age to the precision that 1 small frequency spacing will provide.
It is also interesting that  $\delta \nu$  coupled
with a small radius error had a strong impact on  
the precision of the mixing-length parameter $\alpha$.

\item
We know that a discrepancy exists between the observed
and the model frequencies, which leads to a systematic offset in 
mass determination (e.g. Miglio \& Montalb{\'a}n 2005).
By allowing the radius to determine the mass, we have the advantage that 
1) we have an independent measurement that we can use 
to try to resolve the discrepancy, and 
2) by using the radius to determine mass, we can use the 
frequencies to probe the stellar interior.

\item
One final remark we would like to make is how effective a 
radius measurement can be to detect some error in the model.
During this study we conducted some 
{\it hare and hounds}\footnote{Searching for model parameters when OLC is unaware 
of the real values.} 
tests. 
In one particular case, we found that 
 all of the observations were fit very well
except for the radius, which showed a larger than 3$\epsilon$ deviation value.
This turned out to be the result of an incorrectly quoted 
metallicity meausurement!  This just shows how powerful the radius can be to 
detect an error/flaw, and this could possibly be used to detect a
flaw in the physics of the models.
\end{itemize}

\acknowledgments
 OLC would like to thank
   Andrea Miglio, Tim Bedding and Hans Kjeldsen for some stimulating 
   questions.  The authors are also extremely greatful for the 
   constructive comments and suggestions given by the referee
   Pierre Kervella.
This work was supported in part by a grant (OC) within the project
   {\scriptsize POCI/CTE-AST/57610/2004} from {\sl Funda\c c\~ao 
   para a Ci\^encia e Tecnologia} and POCI 2010 with
   funds from the European programme {\sl Fundo Europeu De 
   Desenvolvimento Regional}.


\begin{table}[h]
\begin{center}
\caption{System Parameters \label{tbl-parameters}}
\begin{tabular}{lr}
\tableline\tableline
Parameter &Value \\
\hline
$M$  (M$_{\odot}$) & 1.030 \\
$\tau$ (Gyr) & 1.00  \\
$X$  &0.740\\
$Z$  & 0.018\\
$\alpha$ & 1.50\\
\hline
\end{tabular}
\end{center}
\end{table}

\begin{table}[h]
\begin{center}
\caption{System Observables \label{tbl-observables}}
\begin{tabular}{lrl}
\tableline\tableline
Measurement &Value & Error \\
& ($O_i$) &($\epsilon_i$)\\
\hline
$R_{\star}$ (R$_{\odot}$) & 0.946 \\
T$_{\rm eff}$ (K) & 5421 &50 \\
$\log g$ &4.5 &0.3\\
$[$M/H$]$ &0.00 &0.05\\
M$_V$ (mag)  &4.50 &0.05\\
(U-V) (mag) & 0.633 &0.005\\
(V-R) (mag) & 0.400 &0.005 \\
$\bar{\Delta \nu}$ ($\mu$Hz)  & 148.3\\
$\bar{\delta \nu}$ ($\mu$Hz) & 14.6\\
$\nu_{0,12}$ ($\mu$Hz) &1996.9 \\
\hline

\multicolumn{3}{l}{\small The errors on $R_{\star}$ and $\nu$ are varied.}
\end{tabular}
\end{center}
\end{table}




\twocolumn

\twocolumn


\end{document}